\documentclass[twocolumn,trackchanges]{aastex631}

\usepackage[version=4]{mhchem}
\usepackage{float,graphicx,multirow}

\begin{document}


\title{Explaining the Chemical Inventory of Orion KL through Machine Learning}

\author[0000-0002-8505-4459]{Haley N. Scolati}
\affiliation{Department of Chemistry, University of Virginia, Charlottesville, VA 22903, USA}
\author[0000-0001-9479-9287]{Anthony J. Remijan}
\affiliation{National Radio Astronomy Observatory, Charlottesville, VA 22903, USA}
\author[0000-0002-4649-2536]{Eric Herbst}
\affiliation{Department of Chemistry, University of Virginia, Charlottesville, VA 22903, USA}
\affiliation{Department of Astronomy, University of Virginia, Charlottesville, VA 22903, USA}
\author[0000-0003-1254-4817]{Brett A. McGuire}
\affiliation{Department of Chemistry, Massachusetts Institute of Technology, Cambridge, MA 02139, USA}
\affiliation{National Radio Astronomy Observatory, Charlottesville, VA 22903, USA}
\author[0000-0002-1903-9242]{Kin Long Kelvin Lee}
\affiliation{Accelerated Computing Systems and Graphics, Intel Corporation, 2111 NE 25th Ave, Hillsboro, OR 97124, USA}

\correspondingauthor{Haley N. Scolati, Kin Long Kelvin Lee}
\email{hns3nh@virginia.edu, kin.long.kelvin.lee@intel.com}

\begin{abstract}

The interplay of the chemistry and physics that exists within astrochemically relevant sources can only be fully appreciated if we can gain a holistic understanding of their chemical inventories. Previous work by \citet{Lee2021} demonstrated the capabilities of simple regression models to reproduce the abundances of the chemical inventory of the Taurus Molecular Cloud 1 (TMC-1), as well as provide abundance predictions for new candidate molecules. It remains to be seen, however, to what degree TMC-1 is a ``unicorn'' in astrochemistry, where the simplicity of its chemistry and physics readily facilitates characterization with simple machine learning models. Here we present an extension in chemical complexity to a heavily studied high-mass star forming region: the Orion Kleinmann-Low (Orion KL) nebula. Unlike TMC-1, Orion KL is composed of several structurally distinct environments that differ chemically and kinematically, wherein the column densities of molecules between these components can have non-linear correlations that cause the unexpected appearance or even lack of likely species in various environments. This proof-of-concept study used similar regression models sampled by Lee et al. to accurately reproduce the column densities from the XCLASS fitting program presented in \citet{Crockett2014}. 

\end{abstract}


\section{Introduction} 
\label{sec:intro}

Chemical models are an invaluable tool for gaining insight into the chemistry and physics occurring within the interstellar medium (ISM). Molecules can serve as tracers to highlight specific processes in astrochemically relevant sources, directly correlating our understanding of these regions and the completeness of their chemical inventories: both detected molecules, and robust upper limits to their abundances. When coupled with new observational detections and experimentally derived reaction rates, chemical models are better constrained to accurately describe the astrophysical processes. However, models heavily rely on understanding the formation chemistry and can be limited in scope in observational sources or molecular families that are not well understood \citep{Remijan2023}. Incorporating newly detected molecules into the networks can be incredibly time consuming as multiple new production and destruction pathways must be added, in addition to updating previous mechanisms. The work required to maintain complex networks is a combinatorial problem, and ultimately not sustainable with respect to the current pace of observational and experimental efforts \citep{McGuire2022}.

\citet{Lee2021} alleviated this bottleneck by coupling supervised and unsupervised machine learning methods with cheminformatics to reproduce the known chemical inventory and predict chemical abundances of unobserved species for the well-studied and characterized source, the Taurus Molecular Cloud 1 (TMC-1). Traditional chemical models require chemical and physical parameters (e.g. reaction rates, gas density) based on assumptions since these values cannot always be readily derived or are unknown. Our machine learning approach does not require this information as input, since the molecular species are encoded as compact vector representations that capture chemical and physical properties implicitly using unsupervised learning, and the abundances of unseen molecules are expressed as simple (non)linear functions based off those that have been observed previously. Thus, characterizing the chemistry of an astrophysical source can potentially be an iterative and self-consistent process of fitting a regressor, predicting unobserved abundances, observing, re-fitting, and so on. While this has been demonstrated in TMC-1, it has not yet been investigated whether machine learning characterization can be successfully applied to more complex sources such as the Orion Kleinmann-Low (Orion KL) nebula, a molecule-rich region notorious for being composed of several structurally and chemically distinct environments.

 The majority of this paper will provide a thorough discussion on the predictive nature of our updates to the methodology outlined in\citet{Lee2021}, its applications for astrochemists and astronomers, and chemical inferences that can be made through mutated counterfactual structures. In Section 2, we provide a description of the dataset, as well as a brief overview of the molecular representations and model pipeline from embedded vectors to column density predictions. In Section 3, we discuss the performance of our algorithm’s ability in characterizing the Orion KL nebula and its regions as well as provide column density predictions for target molecules. Finally, we attempt to add explanation to our predictions using counterfactuals to answer the broader question of why some molecules are more abundant than others.


\section{model pipeline \& workflow} 
\label{sec:comp_methods}

\begin{figure*}
\centering
\includegraphics[width=0.8\textwidth]{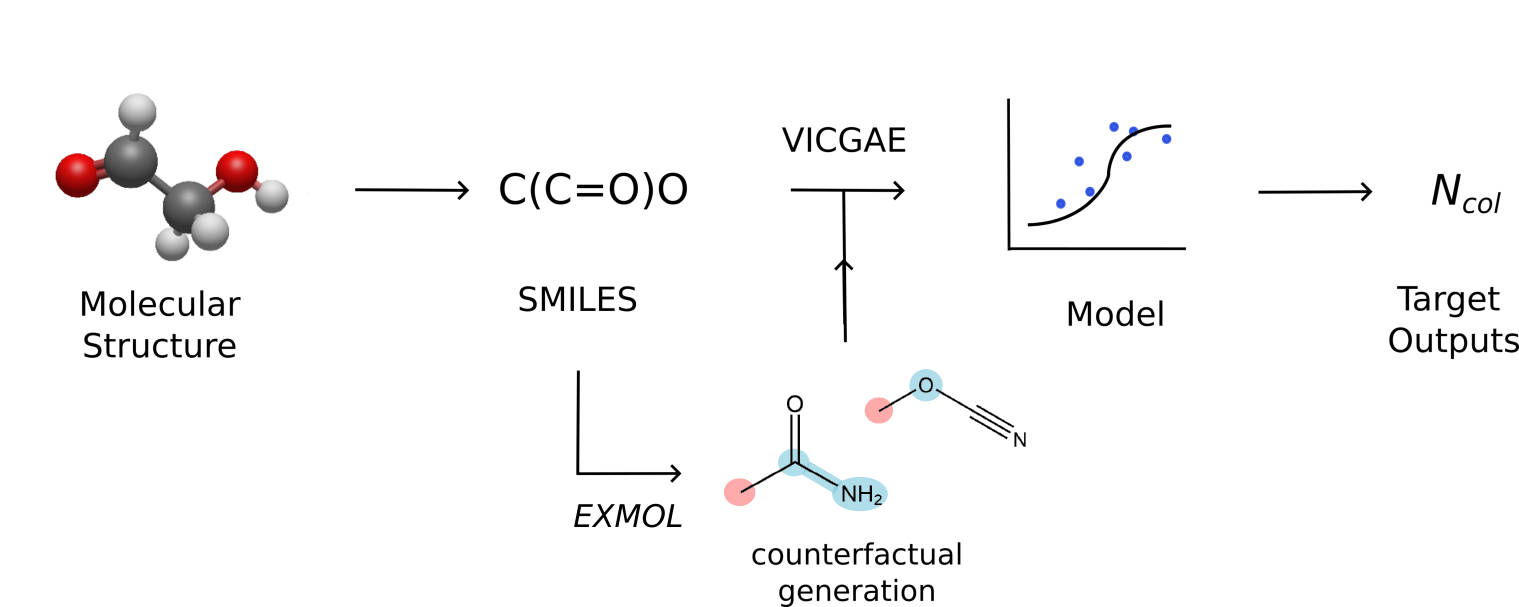}
\caption{Workflow of the supervised machine learning methodology, beginning with molecular structures as SMILES string representations and ending with column density predictions. The strings are encoded into a vector chemical space using the \textsc{VICGAE} embedder. The vectors are then fed into a pretrained model where the target outputs ($N_{col}$) are generated. String representations of counterfactual structures produced with the \textsc{exmol} package \citep{wellawatte_seshadri_white_2021} can also be transformed and provided outputs.}
\label{fig:pipes}
\end{figure*}

The overall model pipeline and workflow from molecular structures to column density predictions is depicted in Figure \ref{fig:pipes}. The initial step in the machine learning (ML) algorithm is to encode the molecular structures and physical parameters into a format that is interpretable by the algorithm. The embedding process allows each structure to be thoroughly described by its molecular features (e.g. connectivity, functional groups, bond lengths) in a vector space. The chemistry incorporated into our model is based solely on the embedded molecular structures. The physically distinct environments of Orion KL requires the physics of the source to be considered and encoded as well. These descriptors were selected from the parameters included by \citet{Crockett2014} in the XCLASS fitting program to local thermodynamic equilibrium (LTE) conditions and only those that assisted the algorithm in differentiating the regions with respect to the physics of the source were chosen. These parameters include the rotational temperature ($T_{rot}$), radial velocity ($v_{lsr}$), and line width ($\Delta{v}$).

\subsection{Molecular representation and embedding} 
\label{subsec:embedder}

Molecular string grammars such as the simplified molecular-input line-entry system (SMILES) format \citep{Weininger1988, OBoyle2012} are compact and efficient ways of representing molecules in a computer and human readable fashion. In the work by \citet{Lee2021}, the authors opted to base their chemical embeddings on the \textsc{mol2vec} algorithm \citep{Jaeger2018}, which adapts natural language models to operate on SMILES rather than using the molecular structures themselves. Without modification, this approach is unable to accommodate for rare isotopologues and vibrationally excited states, both of which are relevant to a holistic description of Orion KL. 

A more recent language for representing molecules---referred to as SELF-referencing embedded strings (SELFIES)---was developed by \citet{Krenn2020} and uses a significantly simplifed syntax and grammar designed especially for machine learning applications. There are many similarities between SMILES and SELFIES, but perhaps the most striking advantage is that every SELFIES string corresponds to a chemically valid molecule: it enables language models to converge more gracefully to generate compact and chemically meaningful vector representations. Appropriately, we developed a new embedding model that adopts SELFIES as the main language, taking advantage of its simplicity and including support for isotopologues.

In the original \textsc{mol2vec} algorithm, conventional fingerprinting approaches are used to isolate common ``building blocks'' to create a corpus (i.e. words within a dictionary) to describe a set of molecules of interest. The embedder used in the current work differs by operating on the SELFIES strings directly in a sequence-to-sequence autoencoder fashion, inspired by the variational autoencoder described in \citet{Krenn2020}. The embedder is trained on the same dataset as described in \citet{Lee2021}, which comprises both terrestrially and astrochemically relevant molecules as a masked language modeling task, and a gated recurrent neural network \citep[GRU,][]{choPropertiesNeuralMachine2014, choLearningPhraseRepresentations2014} learns SELFIES grammar by predicting masked tokens within a sequence. To improve the quality of the learned embeddings, we also include additional regularization terms---the so-called variance-invariance-covariance (VIC) approach \citep{bardesVICRegVarianceInvarianceCovarianceRegularization2022}---that have been shown to instill qualities that are typically obtained through contrastive self-supervised learning tasks, which are significantly more computationally demanding. We refer to this approach as a variance-invariance-covariance gated recurrent autoencoder (VICGAE) model, which can be found in \citet{lee_kin_long_kelvin_2021_7559628}. The VICGAE model is subsequently used in this work to map molecular strings into their corresponding vector representations by processing a given sequence through the encoder GRU. These compact vectors are passed to downstream classical machine learning regressors which we discuss in subsequent sections.

\subsection{Dataset description}

A thorough spectral line survey of Orion KL was done using the HIFI instrument on the \emph{Herschel Space Observatory} that covered 1.2 THz (408--1907 GHz) of bandwidth in the submillimeter region at a resolution of 1.1 MHz. The detection of 39 new molecules, including 79 isotopologues, was reported by \citet{Crockett2014}. These detections were combined with the ground-based detections obtained in the millimeter region using the IRAM 30 m telescope \citep{Tercero2010}. We compiled our dataset using the same molecular catalogs from the above surveys, sourced from the CDMS \citep{Muller2001} and JPL \citep{PICKETT1998} databases and modeled with the XCLASS program. Our full training and testing dataset contains 172 entries of 64 unique molecules, including eighteen isotopologues, nine deuterated, eight vibrationally excited, and nine molecules with multiple velocity components. Each detection of a molecule in an individual region of Orion KL, as provided by \citet{Crockett2014}, are treated as separate data points (i.e. the chemical vector is repeated for the same molecules, but with different physical parameters) to account for chemical overlap between environments as well as cases where a molecule may have multiple velocity components or rotational temperatures. With respect to an astrochemical context, although this number of detections can adequately represent the source’s chemistry, it is small compared to typical ML datasets.

Given that the dataset size greatly affects the performance of ML algorithms, astrochemical applications are at a disadvantage as molecular detections and inventories are generally sparse in number and chemical diversity, respectively. To compensate for the former and mitigate overfitting, we use bootstrapping where data points are sampled and replaced back into the population to create synthetic observations, increasing our effective dataset to 500 observations. We note, however, that regions that are poorly characterized (e.g. hot core south, extended ridge) will only benefit to a certain extent from bootstrapped sampling as the overall chemistry in those regions is still vastly undersampled (i.e. low diversity in the types of molecules detected).

In addition to reproducing the chemical inventory as a whole, we attempted to capture the chemistry of the individual regions. Despite the distinct environments that arise from physical variations within the nebula, there is notable chemical overlap between the environments that further complicates mapping the chemical inventory to the observed abundances \citep{WidicusWeaver2012, Pagani2017}. To differentiate the regions with the ML algorithms, we include additional descriptors that describe the physical attributes of the source. In total, we include six handpicked physical features: three continuous values representing rotational temperature ($T_{rot}$), radial velocity ($v_{lsr}$), and line width ($\Delta{v}$) parameters, and three integer encodings used to indicate vibrational excitation, binary classification of whether the molecule is an isotopologue, and an integer class label (0 through 4) to indicate the cloud environment (e.g. hot core, compact ridge) for each entry. The dataset, including the handpicked physical parameters and environment descriptors, is available for download through Zenodo \citep{hnscolati_2023_7675609}. The full vector encoding used by ML regressors is the concatenation of a chemical vector from VICGAE (representing structure) and a ``physical'' vector that is composed of the six handpicked features detailed above.

With this representation, we are able to predict the abundance of any molecule that can be embedded by VICGAE, given some estimate of its physical conditions, and the model parameterizes---at a high level---\emph{``What} is the column density of a molecule, assuming a set of physical conditions?''

\subsection{Training pipeline and model selection} 
\label{subsec:pipes}

The computational workflow follows the same methodology of that of \citet{Lee2021} where the molecular structures are converted into string representations and encoded to generate a multi-dimensional vector. The vector space consists of the embedded molecular features as well as the physical parameter features. The previous work implemented principal components analysis (PCA) to reduce the dimensionality of the dataset in order to optimize and limit memory usage and computational expense; however, since our dataset is relatively small, the aforementioned had little to no significance and was not performed. 

After preprocessing as described in Section 2.1, the dataset is split into a training and a testing set, then bootstrapped such that the number of data points are increased through random sampling and replacement. The majority (80\%) of the data is used to train the model to map the features to the correct corresponding column densities. During the training process, the model iteratively minimizes a given loss function that calculates that distance between the model and expected outputs. The trained model is subsequently validated with the remaining 20\% of data points to ensure the model has learned its associations rather than ``memorizing'' or overfitting the data.

\subsection{Counterfactual explanations} 
\label{subsec:exmol}

Despite the simple predictive nature of ML algorithms, these models are considered a ``black box'' as they are unable to offer insight as to \emph{why} a particular molecule is predicted at its abundance. To this end, we relied on methodology originally developed by \citet{wellawatte_seshadri_white_2021} to provide counterfactual explanations of molecular drug activity. Where causal relationships provide direct consequences (e.g. ``\ce{X-CN} is high in abundance because \ce{CN} is an abundant precursor), they can be difficult to model and establish, especially owing to the lack of control over astrophysical environments. Instead, counterfactual explanations offer alternative scenarios that affect the outcome of a model---while they do not offer the same \emph{kind} of interpretability as causal relationships, they nonetheless allow comparisons such as ``molecules with \ce{CH3-} would be more abundant as alcohols than aldehydes''. This offers \emph{chemical} explanations for why certain molecules may be more readily observable than others, in addition to plausible alternative targets to search for in surveys.

In our work, we rely on \textsc{exmol} \citep{wellawatte_seshadri_white_2021} for the generation of counterfactuals. The general approach alters a molecule based on its string representation by inducing one or more random valid replacements of atoms/fragments/functional groups within it \citep{Nigam2021}, and reports changes to the predicted abundance as a function of similarity/distance between the base and mutated structure. By making surgical modifications to the original base molecule, we can explore and correlate chemical functionalization with abundance, as parameterized by some black box function \citep{wellawatte_seshadri_white_2021}. One key advantage with this approach is a low barrier to entry: \textsc{exmol} wraps around regressors trained throughout this work without need of modification to the model and/or datasets. We applied this methodology to our Orion KL dataset as a proof of concept to explain the apparent relative abundances of species detected and undetected towards this source.

As previously mentioned, the \textsc{exmol} package was originally developed for drug activity applications, which has a larger and more diverse chemical space compared to that available in astrochemistry. Consequently, without modification or constraint to the ``alphabet'', or library of possible replacements, counterfactuals generated are not astrochemically relevant (e.g. halogenation, organometallic substitutions). The first modification we performed to the procedure was to curate a SELFIES/SMILES alphabet that only contained the atoms, functional groups, and fragments applicable to our dataset---for example, excluding rare earth metals such as scandium. Isotopes and deuterium were also excluded from the alphabet to alleviate an overwhelming number of isotopologue counterfactuals from being generated, however can be trivially included if an appropriate filtering algorithm is applied. Our second modification to the procedure was the similarity/distance metric: in the standard procedure, the Tanimoto distance (the ratio where two vectors intersect in a chemical space) is used: we replaced this metric with cosine similarity (the inner product, or the cosine angle, between vectors in our chemical vector space \citep{Bajusz2015}) based on our embedding vectors. In our experiments, we found that this scoring function provided a significantly more intuitive and representative measure of distance for astrochemically relevant species, in addition to being computationally more efficient.


\section{results and discussion} \label{sec:results_en_disc}
\subsection{Model Performance}

\begin{figure*}
\centering
\includegraphics[width=0.8\textwidth]{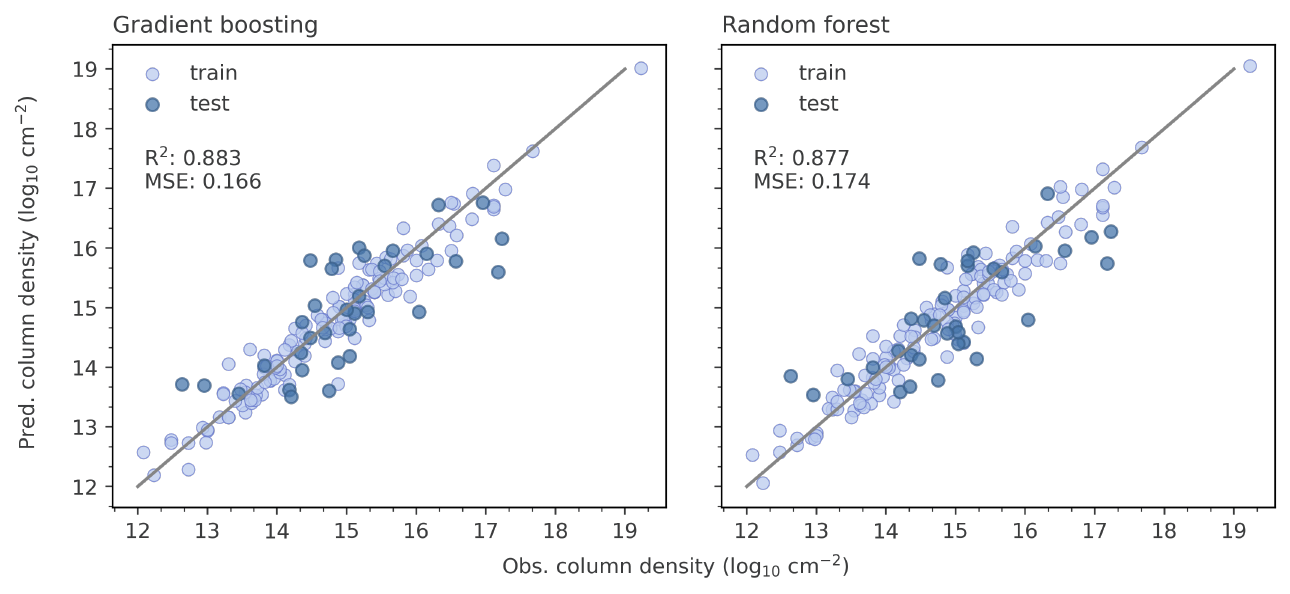}
\caption{Model performance examined by observed column densities plotted against model predictions. The training and testing split of the dataset is indicated by the light blue (training) and dark blue (testing) scatter points. For visualization clarity, only the original data points have been plotted (i.e. bootstrapped data points were omitted from the plots).}
\label{fig:split_all}
\end{figure*}

In this work, the gradient boosting regression (GBR) and random forest regression (RFR) models were used as both are highly efficient ensemble methods: both methodologies were tested in the earlier work by \citet{Lee2021} and provided a good compromise between model complexity/usage and accuracy. Both models were implemented using the \textsc{scikit-learn} package \citep{Pedregosa2011}. The GBR and RFR algorithms were able to accurately reproduce the observed column densities of Orion KL as a whole, despite the limited number of confirmed detections compared to TMC-1. As shown in Figure \ref{fig:split_all}, the performance of both algorithms was comparable, with gradient boosting yielding a slightly better $R^2$ value (0.883) and lower mean squared error (MSE; 0.166) compared to random forest ($R^2$: 0.877, MSE: 0.174). Despite bootstrapping, some overfitting was still present, indicated by the larger spread in the testing set error compared to the training set in Figure \ref{fig:split_all}. Both validated models are available via GitHub (\url{https://github.com/HumbleSituation164/orion-kl-ml}) as well as \citet{haley_n_scolati_2023_7958251}.

\begin{figure*}
\centering
\includegraphics[width=0.8\textwidth]{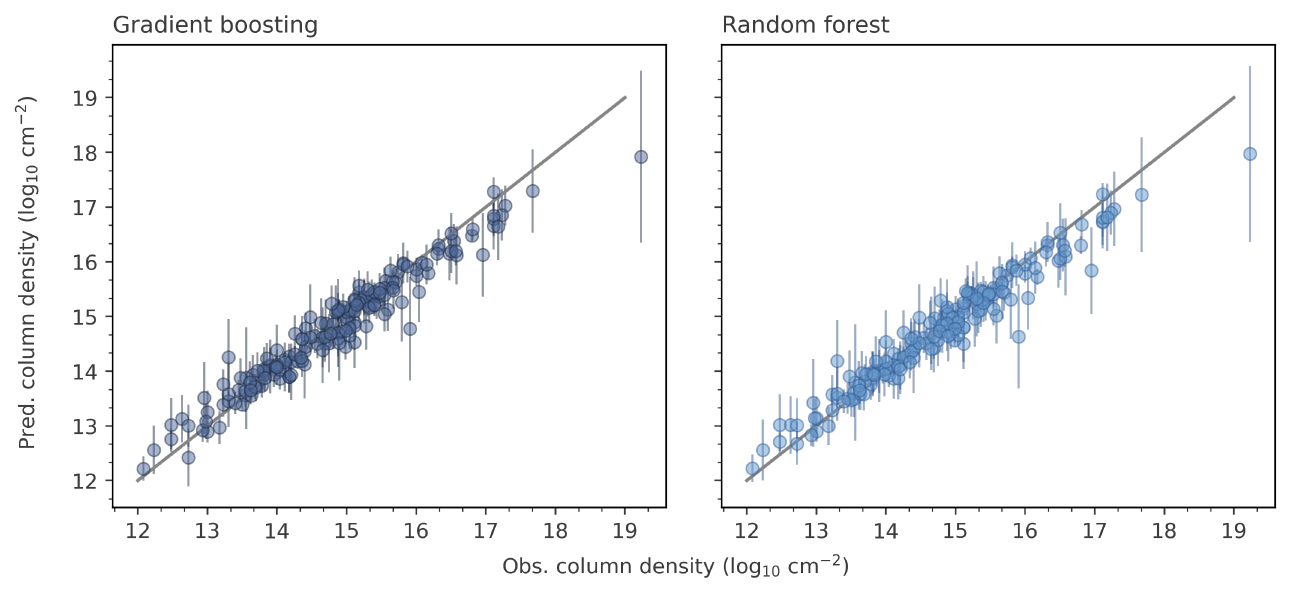}
\caption{Uncertainty estimations for Gradient Boosting and Random Forest regressions were calculated using an ensembling method to calculate the average predicted column density value for each molecule within the dataset. The average predicted value is plotted against the reported observed value. Error bars reflect uncertainty within 1$\sigma$.}
\label{fig:ensemble}
\end{figure*}

To assess uncertainty in the trained models, we used an ensembling approach where multiple of the same model is trained on variations of bootstrapped datasets generated from different random seeds. Predictions were obtained by averaging the results of the ensemble of models, and uncertainty as the standard deviation. Figure \ref{fig:ensemble} provides the uncertainty estimations for the GBR and RFR models where the error bars reflect standard deviation within 1$\sigma$. Although simple, ensembling justifies the level of confidence in the models to reproduce the column densities for this chemical inventory regardless of which set of molecules a model is trained on, and which are used for testing. For molecules with larger standard deviations in their predicted column densities, this is attributed to fewer observations/detections of similar species to allow ensemble members to reach a consensus. We expect that these uncertainties will decrease as additional molecules that share similar chemical composition and structures are detected.

\begin{figure*}
\centering
\includegraphics[width=0.8\textwidth]{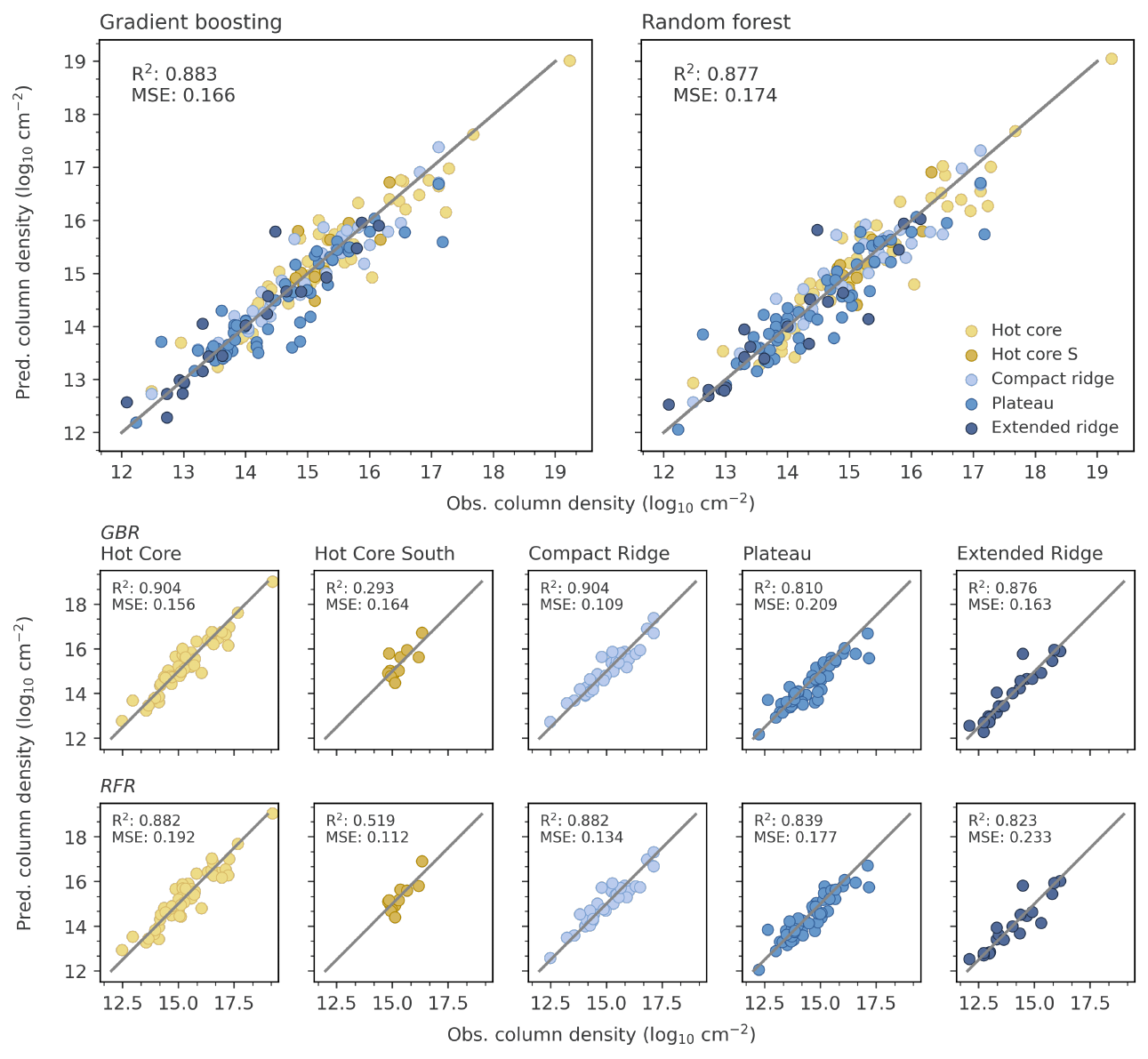}
\caption{Observed column densities plotted against model predictions with respect to each environment within the Orion KL source (top). The individual regions (bottom) are plotted with the same slope = 1 line to indicate linearity in the model performance.}
\label{fig:scatters}
\end{figure*}

In addition to aggregate performance, we assessed how well each model performs with respect to reproducing molecules within each of the respective environments (hot core, hot core south, compact ridge, plateau, and extended ridge). Figure \ref{fig:scatters} correlates the molecular species with their region of detection as well as the regressions for each environment. \ref{tab:detections} in the Appendix provides the list of molecules and their corresponding regions of detection. The environments with more detected species (e.g. hot core and compact ridge) are better reproduced compared to environments with few observed species (e.g. hot core south and extended ridge). For these regions, particularly the extended ridge, the MSE is comparatively larger as a result of the lack of chemical diversity in that environment. As discussed above, without similar structures to train on, the model performance degrades and illustrates the importance in needing a comprehensive inventory if the environment is chemically diverse. However, a comparison of complexity is relative to other sources and ultimately the models are able to reconstruct the Orion KL inventory sufficiently.

\subsection{Column Density Predictions}

\begin{figure*}
\centering
\includegraphics[width=0.8\textwidth]{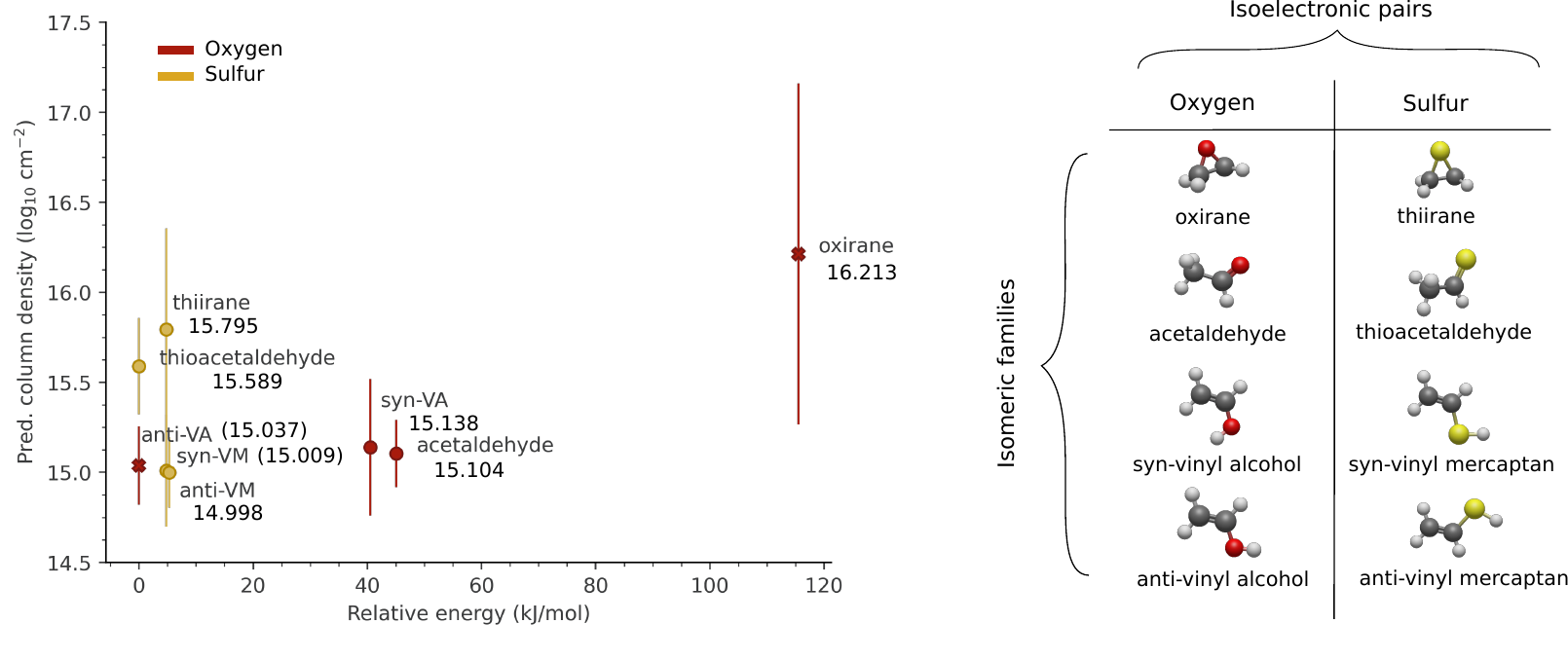}
    \caption{Relative energies of sulfur (thiirane, thioacetaldehyde, syn-/anti-vinyl mercaptan) and oxygen (oxirane, acetaldehyde, syn-/anti-vinyl alcohol)  isomeric families plotted with respect to their predicted column densities (annotated with each molecule). Detected optically thick molecules are indicated with crosses. The corresponding table (right) provides the structures for each molecule, as well as a visualization of the isomeric families and the isoelectronic pairs.}
\label{fig:isomers}
\end{figure*}

The validated GBR and RFR models were applied  to two areas of interest concerning the chemistry of Orion KL: (1) predicting abundances of unobserved molecules and (2) providing baseline predictions for optically thick detections. The physical parameters (rotational temperature, line width, radial velocity) of structurally similar molecules (i.e. functional groups, ring structures, connectivity between atoms) from the original dataset were used as the inputs for the target molecules. The predictions were then made with respect to a specific environment by providing the corresponding numerical environment code as discussed in Section 2.2. Column densities of sulfur and oxygen-substituted isoelectronic families explored in an energetic study by \citet{Martin-Drumel2019} were predicted using the same methodology as previously discussed. Figure \ref{fig:isomers} compares the predicted abundances with the calculated relative energies, where all predictions were made towards the compact ridge. Within the isomeric oxygen family, only the \ce{CH2OCH2} \citep{Tercero2018} and \ce{CH3CHO} \citep{Turner1989, remijan2008complete} isomers have previously been reported in Orion KL. 

According to the minimum energy principle (MEP), the more thermodynamically stable isomers are expected to be higher in abundance \citep{Lattelais2009}. This would suggest for the sulfur isomers to have greater predicted abundances relative to their isoelectronic oxygen analogues; however, Figure \ref{fig:isomers} shows a lack of correlation between the sulfur and oxygen families. Our model predicts \ce{CH2OCH2} to be the most abundant, even though it is also the most energetically unstable. This is indicative that relative abundance is not independent of energetics and further suggests that there is more to consider alongside zero-point bonding energy regarding the potential of a molecule’s detectability \citep{Loomis2015}.

Optically thick detections were not considered in the XCLASS fitting as it was determined by \citet{Crockett2014} that their parameters and models did not provide fruitful physical information. Since our ML model relies solely on the structural composition of the molecules and do not rely on the precise modeling of emission and absorption mechanisms, we are able to set simple baseline expectations for the abundance of molecules shown in Table \ref{tab:opt_thick-pred}, which we hope can be used to constrain chemical network modeling and guide observations.

The approach we have detailed allows for reasonable predictions for species that have yet to be detected, as well as for molecules with low confidence in their derived column densities. These baseline predictions allow for extrapolation for the detectability of unseen molecules and the extent of dynamical effects. The attractiveness of our approach is that general predictions can be made based on the chemical similarity within a vector space. Since our models rely on observations, observational biases must be considered, such as molecules that cannot be detected (i.e. those without permanent dipole moments, etc.). If proxies of these molecules are well constrained by either previous detections or their molecular class is well-characterized within the inventory, they can provide suitable predictions for an unobservable molecule. To put it hypothetically, if two ions have been detected but their parent is unobservable, we can assume the abundances of the ions can allow for a constrained prediction of their parent, which can be useful in chemical modeling.

If a given molecule is predicted poorly, it is likely an indication that observational data pertaining to that particular family of molecules (structurally and chemically) is sparse. As the availability of this data improves, they can be used to iteratively update the dataset and model until predictions become self-consistent.

\begin{table*}[ht]
    \centering
    \caption{Column density predictions (log$_{10}$ cm$^{-2}$) respective to environment for optically thick molecules previously excluded from XCLASS. To keep consistent with Table 3 of \citet{Crockett2014}, ``$\ldots$'' indicates a non-detection and no mark corresponds to an optically thin entry that is already included in their fit. Entries with multiple reported column densities are bold and denote unresolved regions as well as for entries with multiple $T_{rot}$ values.}
    \begin{tabular}{l c c c c}
        \hline
        Molecule & Hot core & Compact ridge & Plateau & Extended ridge \\
        \hline
            \multirow{2}{*}{\ce{CH3CN}}   &   \textbf{15.618}   &   \multirow{2}{*}{15.378}  &  \multirow{2}{*}{15.507}  &   \multirow{2}{*}{...}   \\
                                          &   \textbf{15.662}   &                            &                          &                            \\
        \hline                  
            \ce{CH3OH}     &   16.696     &   16.362  &    ...    &    ...        \\                
            \ce{^{13}CO}   &   15.333     &   14.925  &  14.990   &    14.103    \\              
            \ce{CS}        &   16.033     &   15.811  &  15.268   &    14.944    \\  
            \ce{H2CO}      &   15.822     &   14.908   &          &    ...        \\
        \hline    
            \multirow{2}{*}{\ce{H2C^{18}O}} &  \multirow{2}{*}{15.372}   &  \multirow{2}{*}{14.559}  &  \textbf{14.115}   &  \multirow{2}{*}{...}  \\
                                            &                            &                           &  \textbf{14.921}   &                          \\
        \hline                                    
            \ce{H2S} &  \multirow{2}{*}{15.175}   &  \multirow{2}{*}{...}  &  \textbf{13.945}   &         \\
                     &                            &                        &  \textbf{14.189}   &         \\
        \hline              
            \ce{HCN} &  \textbf{15.931}  &  \multirow{2}{*}{15.902}  &  \textbf{15.489}   &  \multirow{2}{*}{15.275} \\
                    &  \textbf{16.541}  &                            &  \textbf{15.681}   &               \\
        \hline                  
            \ce{HCN, \nu2=1}  &  17.199      &      ...  &      ...     &    ...     \\
        \hline    
            \ce{H^{13}CN}  &  \textbf{14.277}  &       &   \textbf{13.704}  &   \multirow{2}{*}{13.628}   \\
                            &  \textbf{14.495}  &       &   \textbf{13.427}  &           \\
        \hline    
            \ce{HCO+}  &  \multirow{2}{*}{16.017}  &  \multirow{2}{*}{15.377}  &   \textbf{14.804}   &  \multirow{2}{*}{14.344}  \\
                       &                           &                           &   \textbf{15.177}   &                     \\
        \hline    
            \ce{HNC}   &  \textbf{13.688}  &       &   \textbf{13.666}   &   \multirow{2}{*}{13.181}  \\
                       &  \textbf{14.407}  &       &   \textbf{14.589}   &                       \\
        \hline    
            \ce{HNCO}  &  \textbf{14.664}  &    \multirow{2}{*}{...}   &   \textbf{14.489}   &   \multirow{2}{*}{...}   \\
                       &  \textbf{15.330}  &                           &   \textbf{14.659}   &                    \\
        \hline    
            \ce{NH3}          &  16.016    &  \multirow{2}{*}{...}  &   15.042     &     14.317  \\
            \ce{OH}           &     ...    &                      &     14.522     &     ...    \\
        \hline    
            \ce{SiO}  &  ...   &   ...  &   \textbf{14.666}   &   ...     \\
                      &  ...    &  ...   &   \textbf{15.154}   &  ...     \\
                      &  ...    &  ...    &   \textbf{14.835}   &  ...   \\
                      &  ...    &  ...    &   \textbf{15.652}   &  ...    \\       
        \hline                                 
            \ce{SO}   &  \textbf{15.703}   &   \multirow{2}{*}{...}  &   \textbf{15.867}   &            \\
                      &  \textbf{15.842}   &                         &   \textbf{15.858}   &           \\                                 
        \hline    
            \ce{SO2}  &  \textbf{16.257}   &      &  \textbf{15.444}   &         \\
                      &  \textbf{16.565 }  &        &  \textbf{15.901}   &         \\                                 
                      &   ...              &   ...  &  \textbf{15.997}   &     ...    \\     
    \end{tabular}
    \label{tab:opt_thick-pred}
\end{table*}

\subsection{Counterfactual Generation}

\begin{figure*}
\centering
\includegraphics[width=0.8\textwidth]{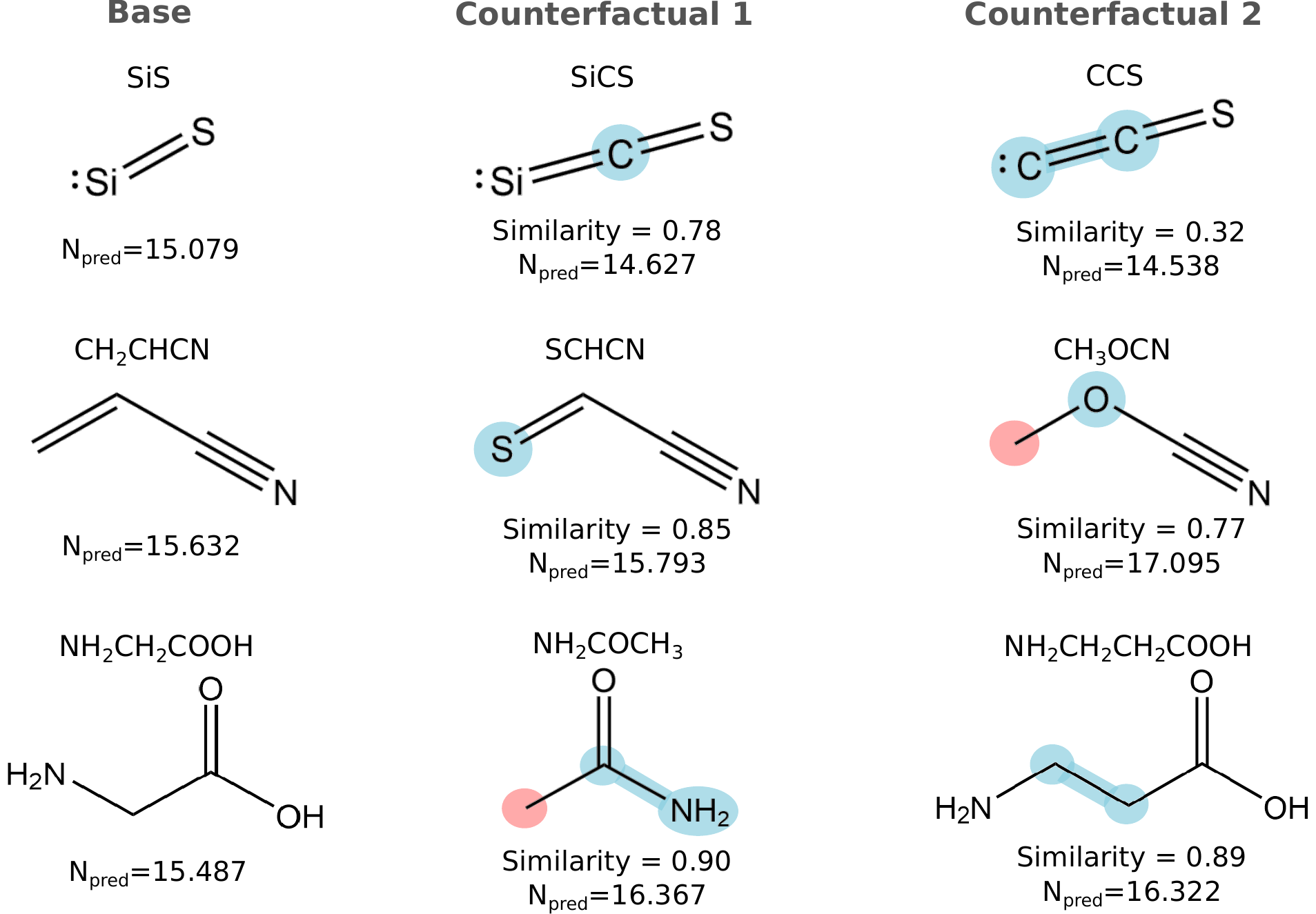}
    \caption{Base molecules and corresponding counterfactual structures generated with our modified \textsc{exmol} procedure. From top to bottom, we provide two counterfactuals (left to right) for the base molecules silicon monosulfide (\ce{SiS}), vinyl cyanide (\ce{CH2CHCN}), and glycine (\ce{NH2CH2COOH}). Additions or ``swaps'' within the molecule are highlighted in blue, while mutations via subtraction are indicated with red. The predicted column densities are reported in units of log$_{10}$ cm$^{-2}$. Cosine similarity scores are also provided for each counterfactual.}
\label{fig:cfs}
\end{figure*}

Figure \ref{fig:cfs} provides two counterfactual examples for each of three base molecules: silicon monosulfide (\ce{SiS}), vinyl cyanide (\ce{CH2CHCN}), and glycine (\ce{NH2CH2COOH}). In the first case, sulfur-bearing silicon species are of interest towards Orion KL owing to their high abundances: with counterfactual explanations, we are able to automate the process of generating similar species to target for future surveys. The most similar counterfactuals generated (excluding \ce{C=S}) are \ce{Si=C=S} and its isoelectronic analogue \ce{C=C=S}, both of which are predicted to be lower in abundance than its base. By extending the linear structure by a single carbon, the predicted column density decreases significantly. This trend is also observed if silicon is replaced with a terminal carbon to give a counterfactual with the same valency, \ce{C=C=S}, which has been detected in four of the Orion KL components \citep{Tercero2010}. 

While the analysis of \ce{Si=S} could be readily done manually, for larger species more substitutions are possible and automation with our modified \textsc{exmol} improves the process significantly. The two other cases, \ce{CH2CHCN} and \ce{NH2CH2COOH}, have previously been searched for towards Orion KL but have not yet been detected and their counterfactuals can be used as a case study to infer preferential formation of related species. \ce{CH2CHCN} (log$_{10}$$N_\mathrm{pred}$=15.632 cm$^{-2}$) is an ubiquitous building block of organic cyanides \citep{Palmer2017} and \ce{NH2CH2COOH} (log$_{10}$$N_\mathrm{pred}$=15.487 cm$^{-2}$) is an amino acid crucial to the origins of life that has been searched for in various sources with little success \citep{Cunningham2007, Snyder2005}. For the former, the model predicts that partial replacement of the vinyl group for \ce{=S}, yielding \ce{SCHCN} (log$_{10}$$N_\mathrm{pred}$=15.793 cm$^{-2}$) only slightly increases the column density, whereas replacing the vinyl group (\ce{-CH=CH2}) with a \ce{CH3-O-R} group to produce \ce{CH3OCN} (log$_{10}$$N_\mathrm{pred}$=17.095 cm$^{-2}$) increases the column density by almost two orders of magnitude. Worded differently, preservation of the double bond (producing \ce{SCHCN}) does not significantly impact abundance and it is likely the motif (\ce{X-CHCN}) that needs to change (i.e. to \ce{CH3OCN}) to improve chances of detectibility. Owing to the large abundance of molecules such as dimethyl ether (\ce{CH3OCH3}), we believe this is a ML-generated hypothesis that could be readily tested in future observing campaigns.

In the case of glycine, which is predicted to have a similar column density (log$_{10}$$N_\mathrm{pred}$=15.487 cm$^{-2}$) to vinyl cyanide, and both counterfactuals increase the column density by nearly an order of magnitude. The first counterfactual replaces the alcohol group (\ce{-OH}) with methyl (\ce{-CH3}) and shortens the internal carbon chain to the terminal amine (\ce{NH3}) to yield acetamide (\ce{NH2COCH3}; log$_{10}$$N_\mathrm{pred}$=16.367 cm$^{-2}$)---the higher abundance could be ascribed to the prevalence of amines, amides, and cyano species (e.g. \ce{NH2D}, \ce{NH2CHO}, and \ce{CH3CN}, respectively) over carboxylic acids towards this source \citep{Crockett2015, Tercero2018}. On the other hand, lengthening the aliphatic backbone by another \ce{CH2} unit is also predicted to increase its column density yielding $\beta$-alanine (\ce{NH2CH2CH2COOH}; log$_{10}$$N_\mathrm{pred}$=16.322 cm$^{-2}$).

For both vinyl cyanide and glycine, our counterfactuals are predicted to be in much higher abundance, which implies that these two molecules are likely in a region of chemical space that are local minima in abundance, leading to difficulties in their detectibility---this observation in itself can be considered as interpretibility in the regressor. We stress that these counterfactual explorations do not establish \emph{causal} relationships, i.e. the \ce{X=CH2CN} group \emph{causes} vinyl cyanide to have a low abundance, but instead correlations. Nonetheless, the counterfactuals provide valuable, viable alternative molecules that could be searched for, both for completeness and for testable hypotheses that could \emph{lead} to causal relationships: for vinyl cyanide, it may be that other \ce{X=CH2CN} substitutions are unlikely to be in high abundance, but cyanic acid (\ce{X-OCN}) derivatives are chemically similar and abundant alternatives to the search for vinyl cyanide. Similar studies could be performed on any arbitrary family of molecules, however complicated by the fact that molecules can elude detection by having unfavorable partition functions, regardless of their abundance. Thus, future work in this avenue may address \emph{detectibility}, rather than abundance, as targets for prediction.

\section{conclusions} \label{sec:the_end}

In this work, we have demonstrated that simple regression models used in \citet{Lee2021} can be successfully extended to more chemically and physically complex astrochemical sources in reproducing their chemical inventories. 

With the SELFIES grammar, we are able to represent additional structural and chemical elements that fully capture the chemistry of Orion KL. As a result, the entire chemical inventory, as well as the individual environments, can be reproduced within a margin of error. One of the key advantages of this approach, as mentioned in \citet{Lee2021}, is that it drastically simplifies the task of predicting molecular properties, such as column density. By relying on representations obtained through unsupervised pre-training, a regressor ML model can be conditioned on existing observations, and predicting abundances of unseen molecules is as straightforward as simply embedding their string representations. This contrasts with conventional chemical models, which require new reactions to be included, their rates estimated, and so on; in the case of isotopologues, an entire separate network must be maintained.

To bridge the predictions with chemical intuition, counterfactual structures were generated using the \textsc{exmol} package to produce ``what-if'' target molecules to infer reasoning as to how the present and non-present structural elements and functional groups affect a molecule’s detectability. The overall appeal in providing baseline predictions allows for another set of constraints to compare with the chemical models. 

The capabilities of our models allows us to make column density predictions directly from string representations of molecules, without the need for prior laboratory and quantum chemical work, or knowledge of the source, and therefore provides a complementary approach to prescreening observation candidates prior to following up with detailed experiments and calculations. This is highly advantageous as the base predictions are well within a magnitude of error and can be made within seconds with a trained model. Here we only considered the molecular entries used in the XCLASS fitting; however, the dataset can be built upon, meaning newly detected molecules and their parameters can be added to further train the models and minimize error. Column density predictions for target molecules can be made quickly and easily---all you need are SMILES.

\begin{acknowledgements}
The National Radio Astronomy Observatory is a facility of the National Science Foundation operated under cooperative agreement by Associated Universities, Inc. H.N.S. acknowledges funding and research support from the National Radio Astronomy Observatory through a Graduate Research Assistantship. E.H. thanks the National Science Foundation (US) for support of his research program in astrochemistry through grant AST19-06489. 
\end{acknowledgements}

\newpage

\bibliography{references}{}
\bibliographystyle{aasjournal}

\appendix

\renewcommand{\thetable}{A\arabic{table}}
\setcounter{table}{0}

\section{Molecular Detections by Region}
\begin{longtable}{ll|ccccc}
\caption{Observed molecules and region(s) of detection towards Orion KL. Detections are indicated with an "X" in the region columns. The dataset was produced based on the inventory of molecules used in the XCLASS fitting as described by \citet{Crockett2014}. The full regression datatset with physical parameters and the train/test split used can be found in \citet{hnscolati_2023_7675609}} \\
\label{dataset}

        Molecule  &  SMILES  &  Hot core  &  Hot core south  &  Compact ridge  &  Plateau  &  Extended ridge \\
    \hline 
        \ce{C2H3CN}             &    C=CC\#N          &    X    &         &         &         &         \\
        \ce{C2H5CN}             &    CCC\#N           &    X    &         &         &         &         \\  
        \ce{CH2NH}              &    C=N              &    X    &         &         &         &    X    \\
        \ce{CH3CN, \nu8=1}      &    CC\#N            &    X    &         &    X    &    X    &         \\
        \ce{^{13}CH3CN}         &    [13CH3]C\#N      &    X    &         &    X    &    X    &         \\   
        $\ce{CH3{^{13}CN}}$     &    C[13C]\#N        &    X    &         &    X    &    X    &         \\
        \ce{C^{18}O}            &    [C-]\#[18O+]     &    X    &         &    X    &    X    &    X    \\
        \ce{C^{17}O}            &    [C-]\#[17O+]     &    X    &         &    X    &    X    &    X    \\
        \ce{H2O, \nu8}          &    O                &    X    &         &         &         &         \\
        $\ce{H2{^{17}}O}$       &    [17OH2]          &    X    &         &    X    &    X    &         \\
        \ce{HDO}                &    [2H]O[H]         &    X    &         &    X    &    X    &         \\
        $\ce{H2{^{34}}S}$       &    [34SH2]          &    X    &         &         &    X    &         \\
        $\ce{H2{^{33}}S}$       &    [33SH2]          &    X    &         &         &    X    &         \\
        \ce{HC3N}               &    C\#CC\#N         &    X    &         &         &    X    &         \\
        \ce{HC3N, \nu7=1}       &    C\#CC\#N         &    X    &         &         &    X    &         \\
        \ce{HCN, \nu2=2}        &    C\#N             &    X    &         &         &         &         \\
        \ce{H^{13}CN, \nu2=1}   &    [13CH]\#N        &    X    &         &    X    &         &         \\
        \ce{HC^{15}N}           &    C\#[15N]         &    X    &         &         &    X    &    X    \\
        \ce{DCN}                &    [2H]C\#N         &    X    &         &    X    &    X    &    X    \\
        \ce{H^{13}CO+}          &    [13CH]=[O+]      &    X    &         &    X    &    X    &    X    \\
        \ce{HNC, \nu2=1}        &    [C-]\#[NH+]      &    X    &         &         &         &         \\
        \ce{HN^{13}C}           &    [13C-]\#[NH+]    &    X    &         &         &    X    &    X    \\
        \ce{HN^{13}CO}          &    O=[13C]=N        &    X    &         &         &    X    &         \\
        \ce{o-NH2}              &    [NH2]            &    X    &         &         &         &    X    \\
        \ce{NH3, \nu2}          &    N                &    X    &         &         &         &         \\
        \ce{^{15}NH3}           &    [15NH3]          &    X    &         &         &    X    &    X    \\
        \ce{NH2D}               &    N[2H]            &    X    &         &         &         &         \\
        \ce{NO}                 &    [N]=O            &    X    &         &    X    &    X    &    X    \\
        \ce{NS}                 &    [N]=S            &    X    &         &    X    &         &         \\
        \ce{OCS}                &    O=C=S            &    X    &         &    X    &    X    &    X    \\
        \ce{^{34}SO}            &    O=[34S]          &    X    &         &         &    X    &         \\
        \ce{^{33}SO}            &    O=[33S]          &    X    &         &         &    X    &         \\
        \ce{SO2, \nu2=1}        &    O=S=O            &    X    &         &         &         &         \\
        \ce{^{34}SO2}           &    O=[34S]=O        &    X    &         &         &    X    &         \\
        \ce{^{33}SO2}           &    O=[33S]=O        &    X    &         &         &    X    &         \\
        \ce{CCH}                &    [C]\#C           &         &    X    &         &         &    X    \\

\\
\\

    (Continued)    \\
    Molecule  &  SMILES  &  Hot core  &  Hot core south  &  Compact ridge  &  Plateau  &  Extended ridge \\
        \hline
        \ce{CH3OCH3}            &    COC              &         &    X    &    X    &         &         \\
        \ce{^13CH3OH}           &    [13CH3]O         &         &    X    &    X    &         &         \\
        \ce{CH3OD-A}            &    CO[2H]           &         &    X    &    X    &         &         \\
        \ce{CH3OD-E}            &    CO[2H]           &         &    X    &    X    &         &         \\
        \ce{CH2DOH}             &    [2H]CO           &         &    X    &    X    &         &         \\
        \ce{^{13}CS}            &    [S+]\#[13C-]     &         &    X    &    X    &    X    &    X    \\
        \ce{C^{34}S}            &    [34S+]\#[C-]     &         &    X    &    X    &    X    &    X    \\
        \ce{C^{33}S}            &    [33S+]\#[C-]     &         &    X    &    X    &    X    &    X    \\
        $\ce{H2{^{13}}CO}$      &    [13C]=O          &         &    X    &    X    &         &         \\
        \ce{H2CS}               &    C=S              &         &    X    &    X    &         &         \\   
        \ce{HD^{18}O}           &    [2H][18OH]       &         &    X    &         &         &         \\
        \ce{D2O}                &    [2H]O[2H]        &         &    X    &         &         &         \\ 
        \ce{C2H5OH}             &    CCO              &         &         &    X    &         &         \\
        \ce{CH3OCHO}            &    O=COC            &         &         &    X    &         &         \\
        \ce{^{13}C^{18}O}       &    [13C-]\#[18O-]   &         &         &    X    &         &    X    \\
        \ce{H2CCO}              &    O=C=C            &         &         &    X    &         &         \\
        \ce{HDCO}               &    [2H]C=O          &         &         &    X    &         &         \\
        \ce{HCS+}               &    C=[S+]           &         &         &    X    &         &         \\
        \ce{HNC}                &    [C-]\#[NH+]      &         &         &    X    &         &         \\
        \ce{NH2CHO}             &    O=CN             &         &         &    X    &         &         \\
        \ce{OD}                 &    [O][2H]          &         &         &    X    &         &         \\
        \ce{SO2}                &    O=S=O            &         &         &    X    &         &    X    \\
        \ce{CN}                 &    [C]\#N           &         &         &         &    X    &    X    \\
        \ce{H2CO}               &    C=O              &         &         &         &    X    &         \\
        \ce{HCl}                &    Cl               &         &         &         &    X    &         \\
        \ce{H^{37}Cl}           &    [37ClH]          &         &         &         &    X    &         \\
        \ce{^{29}SiO}           &    [29Si]=O         &         &         &         &    X    &         \\
        \ce{^{30}SiO}           &    [30Si]=O         &         &         &         &    X    &         \\
        \ce{SiS}                &    [Si]=S           &         &         &         &    X    &         \\
        \ce{H2S}                &    S                &         &         &         &         &    X    \\
        \ce{SO}                 &    S=O              &         &         &         &         &    X    \\

\label{tab:detections}
\end{longtable}

\end{document}